\documentclass[aps,prl,twocolumn,showpacs,amsmath,amssymb]{revtex4}
\usepackage{graphicx}
\usepackage{amsmath}
\usepackage{amsbsy}
\usepackage{dcolumn}
\usepackage{bm}
\newcommand{\w}{{\omega}}
\begin{document}

\title{
Pseudogaps in an Incoherent Metal}

\date{\today}
\author{K. Haule$^{1,2}$, A. Rosch$^{2,3}$, J. Kroha$^{2}$ and P. W\"olfle$^{2}$}
\affiliation{$^{1}$J. Stefan Institute, P.O. Box 3000, 1001 Ljubljana,
Slovenia\\
$^{2}$Institut f\"ur Theorie der Kondensierten Materie, Universit\"at
Karlsruhe, D-76128 Karlsruhe, Germany\\
$^{3}$Sektion Physik, LMU M\"unchen, D-80799 M\"unchen, Germany}

\begin{abstract}
How are the properties of a metal changed by strong
inelastic scattering? We investigate this question within the
two-dimensional t-J model using extended dynamical mean field theory
and a generalized non-crossing approximation.  Short-ranged
antiferromagnetic fluctuations lead to a strongly incoherent single
particle dynamics, large entropy and resistance. Close to the Mott
transition at low hole doping a pseudogap opens, accompanied by a drop
in resistivity and an increase in the Hall constant for both lower
temperatures and doping levels. The behavior obtained bears surprising
similarity to properties of the cuprates.
\end{abstract}
\bigskip
\noindent 

\maketitle 

Most of our present understanding of the properties of metals is based
on Landau's Fermi liquid theory: low-energy excitations are coherent
quasiparticles with the quantum numbers of electrons.  This concept
has been proven to be extremely successful even in systems where
interactions are very strong, e.g. in liquid $^3$He or in heavy
Fermion compounds. In a few classes of materials, however, most
notably the cuprate superconductors, the usual Fermi liquid picture
appears to break down: transport is anomalous, pseudo-gaps open, entropy
is large and various ordering phenomena appear to compete with each
other \cite{andersonBook,arpes,entropie,transport}. This has been
taken to indicate that new low-energy, long wavelength excitations
like spinons and holons or more conventionally spin, charge,
current or pair fluctuations play a dominant role
\cite{andersonBook}. However, a convincing theory based on such
scenarios is still missing.

In this paper, we want to follow a different and less explored route,
investigating the possibility that {\em incoherent} and {\em local}
excitations dominate as it might happen especially at higher
temperatures $T$ when strong quantum and thermal fluctuations driven by
competing interactions decohere the fermionic excitations.  Our
starting point is the two-dimensional t-J model which describes on the
one hand the physics of a doped Mott insulator and on the other hand
the physics of an antiferromagnetic (AF) superexchange interaction
between nearest neighbor spins.  Long range AF order
(in 2d possible only at $T=0$) gets destroyed by a few
percent hole doping.  The resulting spin state is characterized by
short range AF correlations, and highly incoherent excitations, which
are difficult to describe in any conventional many-body scheme relying
on quasi-particle excitations.  The incoherent character of
excitations in the cuprates is clearly seen in the high electrical
resistivity, the large relaxation rates for spin and charge, and the
large entropy.

Prominent feature of the underdoped cuprates is the pseudogap
in the single particle  \cite{arpes} and particle-hole spectra.  A plausible
explanation for it involves the effect of finite ranged fluctuating
antiferromagnetic or superconducting domains leading to a distribution
of local spin gaps \cite{Rosch_Pseudo}.  
We will show below that there
is a different source of pseudogaps arising through nearest neighbor
exchange coupling and retardation effects in the presence of strong
magnetic fluctuations.

Our approximation scheme, based upon the 
extended dynamical
field theory (EDMFT) \cite{Si_PRL}, see below, neglects most of
the longer-range non-local aspects of the problem but include the
strong inelastic scattering of electrons from local magnetic
fluctuations. By comparing our results to experiments on the cuprates
and to numerical results for the t-J model, we investigate 
 to what extent features like the pseudogap, the large
entropy or the Hall effect can be described by a strongly incoherent
metal.

{\it Model:}
The t-J model describes electrons in a tight-binding model subject to
(i) the constraint of at most singly occupied lattice sites (effected
by projected fermion creation and annihilation operators, 
$\tilde{c}_{i\sigma}^+={c}_{i\sigma}^+ (1-n_{-\sigma})$), 
and (ii) to an AF
spin interaction,

\begin{equation}
H=-\sum_{i,j;\sigma} t_{ij}\tilde c_{i\sigma}^+\tilde c_{j\sigma} +
\frac{1}{2} \sum_{i,j} J_{ij} \vec S_i\cdot \vec S_j
\label{1}
\end{equation}
where $\vec S_i = \frac{1}{2} \sum_{\sigma,\sigma{'}} \tilde
c_{i\sigma}^+\vec\tau_{\sigma\sigma{'}} \tilde c_{i\sigma{'}}$ is the
spin operator at lattice site $i$, $\vec\tau$ denoting the vector of
the Pauli matrices, and $t$ and $J$ couple only nearest neighbors.

As our goal is to describe the high-$T$ incoherent regime, we
will neglect most spatial correlations, assuming that the self-energy
of the electrons is local, $\Sigma_{\vec{k}}(\w)=\Sigma(\w)$. At the
same time, we will keep track of all $\w$-dependences as we consider a
situation where inelastic scattering is very strong. We therefore use
the so-called ``dynamical mean field theory'' (DMFT)
\cite{Metzner_PRL,Kotliar_review}. Taken as a purely local
approximation, DMFT neglects the intersite $J$ term, an important
source for inelastic scattering. To include its effect, we consider
the so-called ``extended'' DMFT (EDMFT) proposed in \cite{Si_PRL}.
This approximation is probably best visualized \cite{Kotliar_review}
by selecting a single site, the ``impurity'', out of the lattice. As
we want to neglect spatial correlations, we can treat the surroundings
as an effective medium providing a fluctuating environment which
consists both of electrons and bosonic spin fluctuations due to
the coupling by $t$ and $J$, respectively.  Local
correlation functions can therefore be calculated by solving the
following quantum impurity model:
\begin{eqnarray}
H_{\text{imp}} &= \sum_{\vec{k}\sigma}E_k c_{\vec{k}\sigma}^+ 
c_{\vec{k}\sigma}
+V \sum_{\vec{k}\sigma} (c_{\vec{k}\sigma}^+ \tilde d_\sigma +
h.c.)\nonumber \\
& - \mu n_d + \sum_{\vec{q}}\omega_q \vec{h}_{\vec{q}}^+
 \cdot \vec{h}_{\vec{q}} + I
\sum_{\vec q} \vec{S}_d (\vec{h}_{\vec{q}} + \vec{h}_{-\vec{q}}^+)
\label{3}
\end{eqnarray}
Here $ \tilde{d}_\sigma^+$ is a projected fermion creation operator
for the impurity orbital, $n_d = \Sigma_\sigma \tilde d_\sigma^+\tilde
d_\sigma$ and $\vec{S}_d = \frac{1}{2} \Sigma_{\sigma, \sigma{'}}
\tilde d_\sigma^+\vec{\tau}_{\sigma,\sigma{'}}\tilde d_{\sigma{'}}$.
The (unrestricted) fermion operators $c_{\vec{k}\sigma}^+$ create a
fermionic bath, the boson operators $\vec{h}_{\vec{q}}^+$ create a
bosonic spin bath (local magnetic field) coupling to the impurity
degrees of freedom.  The effective medium, characterized by the
fermion and boson energies $E_{\vec{k}}$ and $\omega_{\vec{q}}$ has
to be determined self-consistently by identifying the single particle
Greens function and spin susceptibility of the impurity model with the
local Greens function $G_{00}$ and local susceptibility $\chi_{00}$ of
the lattice model
\begin{eqnarray}
G_{00}=\sum_{\vec{k}}G_{\vec{k}} (i\omega) & =& \left[i\omega + \mu -
\sum_{\vec{k}} \frac{V^2}{i\omega - E_k} - 
\Sigma (i\omega)\right]^{-1}\nonumber \\
\chi_{00}=\sum_{\vec{q}} \chi_{\vec{q}}(i\omega) & =& \left[\sum_{\vec{q}}
\frac{2 I^2\omega_q}{(i\omega)^2 - \omega_q^2}\ + \chi_{ir}^{-1}
(i\omega)\right]^{-1}
\label{4}
\end{eqnarray}
Here we use  that within EDMFT the Green's function $G_{\vec{k}}$
and the spin susceptibility 
$\chi_{\vec{k}}$ have a simple $\vec{k}$
dependence as both the self-energy $\Sigma(i\omega)$ and the
irreducible susceptibility $\chi_{ir}(i\omega)$ are taken to be
independent of  $\vec{k}$
\begin{eqnarray}
G_{\vec{k}\sigma} (i\omega)  = \frac{1}{i\omega + \mu - \epsilon_k -
\Sigma(i\omega)}, \, \,  \chi_{\vec{q}}(i\omega)  = \frac{1}{\chi_{ir}^{-1}(i\omega) +
J_q}  \nonumber
\end{eqnarray}
where $\epsilon_k$ and $J_q$ are the lattice Fourier transforms of
$t_{ij}$ and $J_{ij}$, respectively. 
It follows from (\ref{4}) that only the densities of states $N(\omega)
= \Sigma_{\vec{k}}\delta (\omega - E_k)$ and $D(\omega) = \Sigma_{\vec{q}} [\delta (\omega - \omega_q) - \delta(\omega + \omega_q)]$ are
needed and $E_k$ and $\omega_q$ may be assumed to be isotropic in
momentum space.  Formally EDMFT is exact in the limit of infinite
dimension, $d\to \infty$, if both $t$ and $J$ are scaled proportional
to $1/\sqrt{d}$ \cite{Si_PRL}.

\begin{figure}
\begin{center}
\includegraphics[height=3.3cm]{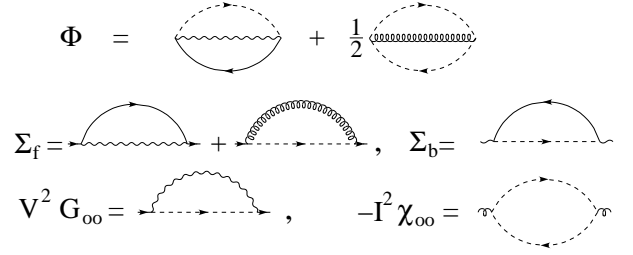}
\end{center}
\caption{\label{figDiagram}The two lowest order contributions to the
  Luttinger-Ward functional $\Phi$ and corresponding self-energies. Only
  diagrams with no line-crossings are taken into account (a generalization 
  of NCA). The broken (wavy) line denotes pseudo fermion
  (pseudo-boson) Green's function $G_{f\sigma}$ ($G_b$), and the solid
  lines represent the conduction electron Green's functions
  $G_{c\sigma}$, the curly line the correlator $G_{h\alpha}$ of the
  bosonic bath. Also shown are the self-energies, the local electron
  Green's function $G_{00}$ and the local susceptibility $\chi_{00}$.}
\end{figure}

The solution of the impurity problem (\ref{3}) for given $N(\omega)$
and $D(\omega)$ is difficult.  Even for the model without the spin
boson field $\vec{h}_{\vec{q}}$ (the well-known Anderson impurity model)
dynamical properties can only be calculated numerically, e.g. using
quantum Monte Carlo, the numerical renormalization group (NRG), or
resummations of perturbation theory like the non-crossing
approximation (NCA) or the conserving T-matrix approximation (CTMA)
\cite{ctma}. Unfortunately,
all of these methods except for the resummation of perturbation theory
and the quantum Monte Carlo method are not easily generalized to
include the spin boson field $\vec{h}_{\vec{q}}$.

We will therefore employ a conserving approximation in which infinite
classes of Feynman diagrams 
are resummed.  We
are aiming at a level of approximation corresponding to NCA for the
usual Anderson model.  In order to effect the projection onto the
sector of Hilbert space without double occupancy of the local energy
level we use 
a pseudo-particle representation,
where the singly occupied state is created by pseudo-fermion operators
$f_{\sigma}^+, \sigma = \uparrow, \downarrow$, whereas the empty
orbital is created by a boson operator $b^+$.  Since the local level
is either empty or singly occupied, the operator constraint $Q = b^+b
+ \Sigma_\sigma f_\sigma^+ f_\sigma = 1$ has to be satisfied at all
times, which can be enforced by adding a term $\lambda Q$ to the
Hamiltonian (\ref{3}) and taking the limit $\lambda \rightarrow
\infty$.  The projected local electron operators in (\ref{3}) may then
be replaced by $\tilde d_\sigma = b^+ f_\sigma$, turning the problem
into a many-body system of pseudo-fermions $f_\sigma$ and slave bosons
$b$, interacting with the fermions $c_{\vec{k}\sigma}$ and bosons 
$\vec{h}_{\vec{q}}$ of the bath.

It is essential that in any approximation one stays within the physical
Hilbert space and does not violate the constraint. Therefore, we
employ a ``conserving approximation'' specified by a generating
Luttinger-Ward type functional $\Phi$ from which all self-energies and
correlation functions are obtained.  We employ the
simplest
conserving approximation by considering only the lowest order diagrams
in $V$ and $I$ (see Fig.~\ref{figDiagram}) using that the effective
hybridization $V$ and the exchange field $I$ are small compared to the
bandwidth $8 t$. Within our conserving approximation, both the local
Green's function of the physical electron and the local susceptibility
can be calculated as a simple convolution of pseudo-particle Green's
functions without vertex corrections (Fig.~\ref{figDiagram}).

\begin{figure}
\begin{center}
\includegraphics[width=0.9 \linewidth]{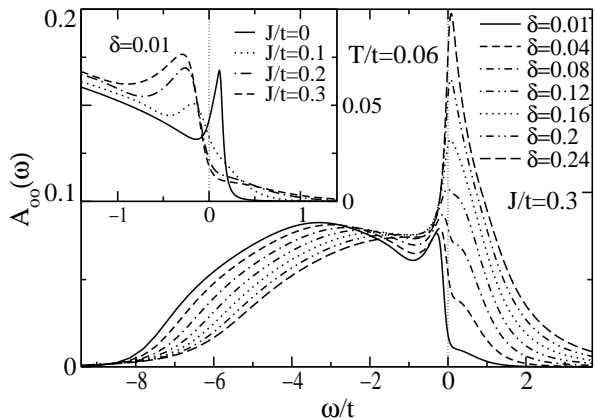}
\end{center}
\caption{\label{figSpectral} The local spectral function plotted
versus frequency for $T=0.06t$ and $J/t=0.3$ for various hole-doping
concentrations $\delta$. Inset: The local spectral function for four
different $J/t=$ 0, 0.1, 0.2 and 0.3 and $T=0.06t$ for a doping of
$\delta=0.01$. The evolution of a pseudogap of width $J$ is clearly
visible.}
\end{figure}

It is important to note that our approximation scheme does not include
the vertex corrections needed to describe correctly how the effective
interactions with the bosonic and fermionic bath renormalize each
other. It therefore cannot be expected to capture correctly the
behavior at low $T$ especially  close to the quantum
critical point where AF order is destroyed by  doping
\cite{Si_Nature}. We believe, however, that in the incoherent high-$T$
regime which is the focus of our study it is unlikely that such vertex
corrections change the physics qualitatively.  On a Bethe lattice, our
EDMFT equations surprisingly are identical to those of a t-J model
with fully {\em random} $J$, as studied within a systematic large $M$
expansion by Parcollet and Georges \cite{olivier}, who did not find
any pseudogaps.  We believe this to be an artifact of their
approximation which uses a Bose-condensed slave boson $\langle
b \rangle$ therefore missing the incoherent part of the spectral
function.

{\it Results:} What happens when inelastic scattering is increased by
switching on a finite $J$? The effect is strongest at small doping as
shown in the inset of Fig.~\ref{figSpectral}: Spectral weight is
pushed below the Fermi energy $E_F$ and a well pronounced pseudogap of
width $J$ opens.  At $T=0.06 t$ the pseudogap closes for $\delta \sim
10\%$ as shown in Fig.~\ref{figSpectral}.

It is tempting to compare our results to experiments in the pseudogap
phase of the cuprates. One should however keep in mind that in the
cuprates nonlocal effects do play an important role, as is evident
from the momentum dependence of the pseudogap. Furthermore, it is
important to stress that we do {\em not} see a pseudogap in the
susceptibility. However, it is interesting to ask which qualitative
features can be understood as a purely local effect. For example, our
approximation scheme explicitly {\em excludes} the possibility that
the reduction of the density of states at $E_F$ is created by the
adjustment of the electronic wavefunction to some small magnetic or
superconducting domains.  The observation that a pseudogap can arise
in an incoherent metal with purely {\em local} correlations is one of
the main results of this paper.  In our approach, the pseudogap opens
when the renormalized chemical potential $\mu-\text{Re} \Sigma(\w=0)$
is pushed towards the edge and finally out of the lower Hubbard band
\cite{haulePhD} by strong magnetic fluctuations: this is only possible
in an incoherent metal when $\text{Im} \Sigma$ is sufficiently large.

\begin{figure}
\begin{center}
\includegraphics[width=0.85\linewidth]{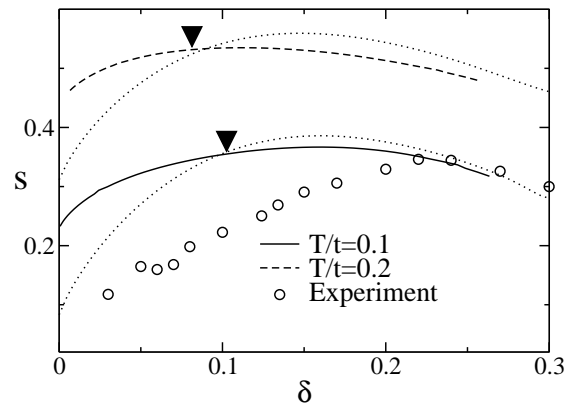}
\end{center}
\caption{\label{figS} Entropy as a function of doping for $J=0.3 t$, $T=0.1 t$
and $T=0.2t$ compared to results from exact diagonalization (dotted
lines) \cite{Prelovsek_ED} and experiments in LSCO \cite{entropie} at
$T \sim 0.07 t$. The triangles mark the doping below which a pseudogap
starts to open in the spectral function.}
\end{figure}

How does this physics manifest itself in other physical quantities?
We calculate the entropy as a crude measure for the relevance of
incoherent excitation from the  free energy $\Omega$: 
\begin{eqnarray}
\Omega/N & = \Omega_{imp} +
\frac{1}{\beta}\sum_{i\omega,\sigma}\sum_{\vec{k}}\ln
\Big[G_{\vec{k}} (i\omega)/G_{00}(i\omega)\Big] \nonumber \\
& - \frac{1}{2} \frac{1}{\beta}\sum_{i\omega,\alpha}\sum_{\vec{q}}
\ln\Big[\chi_{\vec{q}}^{\alpha\alpha}(i\omega)/\chi_{00}^{\alpha\alpha}(i\omega)\Big]
\label{}
\end{eqnarray}
where the impurity contribution in terms of the pseudo-particle
spectral functions $A_{f,b}$ is given by
$e^{-\beta\Omega_{imp}} = \int d\omega
e^{-\beta\omega}\Big[\sum_\sigma A_{f\sigma}(\omega) + A_b(\omega)\Big]$.
The entropy $S =-\partial\Omega/\partial T$ as a function of doping
for various $T$ is shown in Fig. \ref{figS}. First of all,
one realizes that it is rather large even at the lowest temperature of
$T/t = 0.1$, an indication for strong correlations and a rather
incoherent state. The overall magnitude of $S$ compares surprisingly
well with both exact diagonalization and experiments in LSCO (see
Fig.~\ref{figS}). Furthermore, it follows the general trend that
entropy is reduced  both for large doping where the system should
become more coherent and at low doping where magnetic fluctuations
quench the $\ln 2$ entropy of a magnetically disordered Mott insulator
(for $J=0$ the entropy increases for $\delta \to 0$). Interestingly,
the drop in entropy towards low doping occurs precisely when the
pseudogap starts to open in the spectral function (note that in the
experiment both the opening of the pseudogap and the drop in entropy
occur at higher doping).

\begin{figure}
\begin{center}
\includegraphics[width=0.9\linewidth]{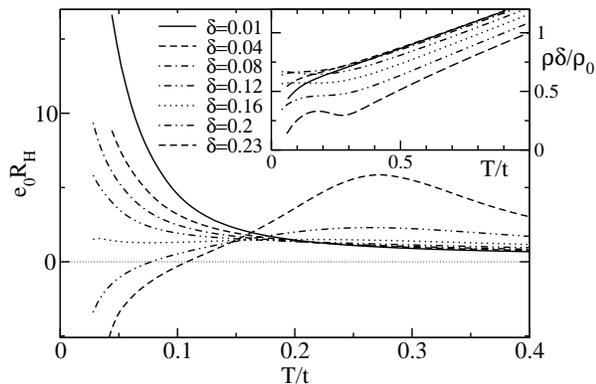}
\end{center}
\caption{\label{figHallT} $T$-dependence of $R_H$ for $J=0.3 t$. For small
doping and $T\to 0$, $R_H$ approaches the value $1/(e_0 \delta)$ expected
for a single hole in a t-J model \cite{Prelovsek_Rh}. Inset:  $T$-dependence of the resistivity multiplied
 by doping $\delta$. The linear $T$ behavior for high $T$ flattens for
 $\delta>0.1$ for $T$ of the order of $J$. For $\delta<0.1$
 the resistivity drops in the regime where a pseudogap opens.}
\end{figure}

How is transport affected by the pseudogap? 
Within EDMFT, there
are no vertex corrections to the conductivities $\sigma_{xx}$ and
$\sigma_{xy}$ which can therefore be directly calculated from the
spectral functions \cite{haulePhD,ekke}.  
In the inset of Fig.~\ref{figHallT} the
$T$-dependence of the resistivity is shown for various dopings. For
high $T$, $\rho$ depends linearly on $T$, an effect which is not related
to the coupling to the bosonic environment as it is also seen in DMFT
\cite{Kotliar_review}. For small doping, the resistivity is
proportional to $1/\delta$: only the holes doped into the Mott
insulator can transport charge.  At the scale of $J$ the resistivity
saturates, probably due to the strong inelastic scattering from spin
fluctuations. Such a behavior is {\em not} observed in
experiments in the cuprates, possibly an indication that non-local
effects and vertex corrections are important for transport.  Note,
however, that in the regime where the pseudogap forms, i.e. for
$\delta<0.1$, the resistivity actually shows a clear drop which is
reminiscent of what is seen experimentally \cite{transport}. 

In Fig.  \ref{figHallT} the $T$-dependence of the Hall constant $R_H$
is displayed for various dopings. We find a strongly growing positive
$R_H$ with decreasing temperature for small dopings and an almost flat
variation for moderate dopings.  In the limit of small doping and low
$T$ the universal relation $R_H = \frac{1}{e\delta}$ is
approached \cite{Prelovsek_Rh}, an indication that Luttinger's theorem
is not applicable in this incoherent regime which cannot be described
by moderately excited Fermi quasiparticles.  Note that the rise of
$R_H$ towards low $T$ seems to happen in the regime where the
pseudogap opens -- in underdoped cuprates a strong increase of $R_H$
with falling temperature is observed upon entering the pseudogap
regime \cite{transport} (the drop of $R_H$ close to $T_c$ is obviously
not included in our theory). In the absence of the coupling to the
bosonic bath, i.e. within DMFT, both the pseudogap
\cite{Kotliar_review} and such an upturn \cite{ekke} are absent.

In conclusion, we have investigated the properties of a highly
incoherent metal close to a Mott insulator subject to strong magnetic
fluctuation. Even purely local magnetic fluctuations change the
physics qualitatively at small doping: they suppress the entropy and
induce a pseudogap by driving the chemical potential out of the lower
Hubbard band. This leads to an increase of the Hall constant and a
drop in the resistivity.  These features are reminiscent of the
behavior seen in the pseudogap phase of the cuprates.  This might
indicate that some of the physics in the cuprates could reflect
properties of a highly incoherent metal with dominating local
fluctuations.  An interesting open question is to what extent
properties of such an incoherent metal are universal and independent
of the details of inelastic scattering mechanisms.

We acknowledge helpful discussions with E.~Abrahams, J.~Bon\v ca, 
A.~Georges, G.~Kotliar, O.~Parcollet,  Q.~Si and especially 
P.~Prelov\v sek. 
Part of this work was supported by the Ministry of Education,
Science and Sport of Slovenia, FERLIN (K.H.) and the Emmy-Noether
program of the Deutsche Forschungsgemeinschaft (A.R.).

\end{document}